\def\year{2022}\relax
\documentclass[letterpaper]{article} 
\usepackage{aaai22}  
\usepackage{times}  
\usepackage{helvet}  
\usepackage{courier}  
\usepackage[hyphens]{url}  
\usepackage{graphicx} 
\usepackage{natbib}  
\usepackage{caption} 
\DeclareCaptionStyle{ruled}{labelfont=normalfont,labelsep=colon,strut=off} 
\usepackage{algorithm}
\usepackage{algorithmic}

\usepackage{newfloat}
\usepackage{listings}
\floatstyle{ruled}
\newfloat{listing}{tb}{lst}{}
\floatname{listing}{Listing}

\usepackage{xcolor}

\usepackage{subfig}
\title{Understanding and Shifting Preferences for Battery Electric Vehicles}
\author {
Nikos Ar\'echiga\thanks{Equal contribution.},
Francine Chen\footnotemark[1],
Rumen Iliev,
Emily Sumner,
Scott Carter,
Alex Filipowicz,
Nayeli Bravo,
Monica Van,
Kate Glazko,
Kalani Murakami,
Laurent Denoue,
Candice Hogan,
Katharine Sieck,
Charlene Wu,
Kent Lyons
}
\affiliations{
    Toyota Research Institute\\
    Los Altos, CA\\
    \{nikos.arechiga, francine.chen, rumen.iliev, emily.sumner, scott.carter,
    alex.filipowicz,
    nayeli.bravo.ctr,
    monica.van.ctr,
    kate.glazko.ctr,
    kalani.murakami.ctr,
    laurent.denoue.ctr,
    candice.hogan,
    kate.sieck,
    charlene.wu,
    kent.lyons\}@tri.global 
}

\begin{document}

\maketitle

\begin{abstract}
Identifying personalized interventions for an individual is an important task. Recent work \cite{harinen2021machine} has shown that interventions that do not consider the demographic background of individual consumers can, in fact, produce the reverse effect, strengthening opposition to electric vehicles. In this work, we focus on methods for personalizing interventions based on an individual’s demographics to shift the preferences of consumers to be more positive towards Battery Electric Vehicles (BEVs). One of the constraints in building models to suggest interventions for shifting preferences is that  each intervention can influence the effectiveness of later interventions. This, in turn, requires many subjects to evaluate effectiveness of each possible intervention. To address this, we propose to identify personalized factors influencing BEV adoption, such as barriers and motivators. We present a method for predicting these factors and show that the performance is better than always predicting the most frequent factors. We then present a Reinforcement Learning (RL) model that learns the most effective interventions, and compare the number of subjects required for each approach.
\end{abstract}

\section{Introduction}

Studies have found that while many people think positively about 
Battery Electric Vehicles (BEV) and would be interested in buying one in the future, they have many uncertainties and unknowns that need to be addressed before buying one for themselves (e.g., \cite{preston2020}). There are a variety of concerns and beliefs about BEV ownership that contribute to this uncertainty, such as ownership cost, limited range, availability of charging, and social and safety factors. These concerns and beliefs can be due to misperceptions about BEVs or not being aware of facts. Some  people who live in a house may believe that charging is inconvenient, and they may not know that BEVs less expensive to maintain than internal combustion engine (ICE) vehicles. Because people have a variety of concerns, and there are many possible interventions to address the range of concerns, interventions will vary in their suitability and effectiveness for a particular individual. When there are many interventions that could be presented, it is important to present the most relevant interventions to a person, rather than showing a fixed set of interventions which may not address their concerns.

In this paper, we examine methods for identifying persuasive interventions for each individual given their background. We use demographic information about an individual to provide a background context. We gather this information through a survey/questionnaire. One of the difficulties in predictions based on survey data from human subjects is the limited data set size for training and testing a model. 

We investigate two primary approaches. The first integrates human knowledge with supervised machine learning for predicting barriers to reduce the number of subjects needed for training a model. Interventions to address the identified barriers can then be presented to a subject. For this approach, the experiments are performed on a dataset collected from the survey. 
In the second approach, reinforcement learning (RL) is used to directly learn which interventions are most effective. These experiments examine the case when no demographic information is available and a limited set of demographic information is available.

\section{Related Work}
The related work falls into several areas: (1) behavior change (2) studies on consumer preferences towards BEVs (3) prediction models with small training sets.
\subsubsection{Behavior Change} 

There has been an increasing acknowledgement of the important role that behavioral change will play for combating environmental challenges \cite{bujold2020}. Techniques and methods which have been developed for improving physical health, mental well-being, educational achievements or productivity levels are currently adapted and applied to resource conservation and climate action \citep{williamson2018climate}. While there are multiple approaches to behavior changes, in this paper we focus on attitudes, which have been considered by many as the core component which behavior change interventions should target \citep{petty2018attitudes}. A popular distinction between different types of interventions is between affective and cognitive interventions \citep{heimlich2008understanding}. Here we are focusing on cognitive interventions alone. 

\subsubsection{Consumer Preferences towards BEVs}
BEVs are seen as a major part of the global effort toward carbon neutrality. When combined with clean energy sources, BEVs will lead to a significant reduction in greenhouse emissions. Various legislative programs in the EU, USA, UK and other countries will be limiting or stopping the sales of non-BEVs in the next decade. The popularity of BEVs, however, is still low, and although consumers are generally interested in BEVs, they still perceive multiple challenges and barriers on the path towards owning one \citep{giansoldati2020barriers}.  
\subsubsection{Small Training Sets} With human subject survey data, it can be challenging to collect a large data set for training a prediction model based on some of the deep learning models. Although a number of techniques, such as transfer learning (e.g., \cite{zhuang2020xferLearning}), fine-tuning (e.g., \cite{tajbakhsh2016fineTuning}), and zero-shot or few-shot learning \cite{snell2017prototypical}, have been developed to address creating models for with new datasets with very few, if any, labels, these techniques make use of an existing large labeled dataset where the input is similar to the input for the target task, e.g., images of objects represented as pixel values. In contrast, human subject data varies widely depending on the question being studied. 
We chose to instead examine models which have been shown to be more efficient to train than deep learning models.

\section{Predicting Barriers to Adoption}
The effect of an intervention in shifting a person’s preferences, such as towards BEVs or PHEVs, can be assessed by either asking a person directly about their preference or showing them a task to indirectly assess it, such as building a car on a website and observing whether they choose to build a BEV, PHEV, hybrid or ICE vehicle. Each presented intervention can affect a person’s preferences, but the effect of an intervention tends to decrease with each presented intervention. Thus it can be difficult to run enough subjects to assess each intervention for the different backgrounds that can influence the effect of an intervention.

\begin{figure}
    \centering
    \includegraphics[width=\columnwidth]{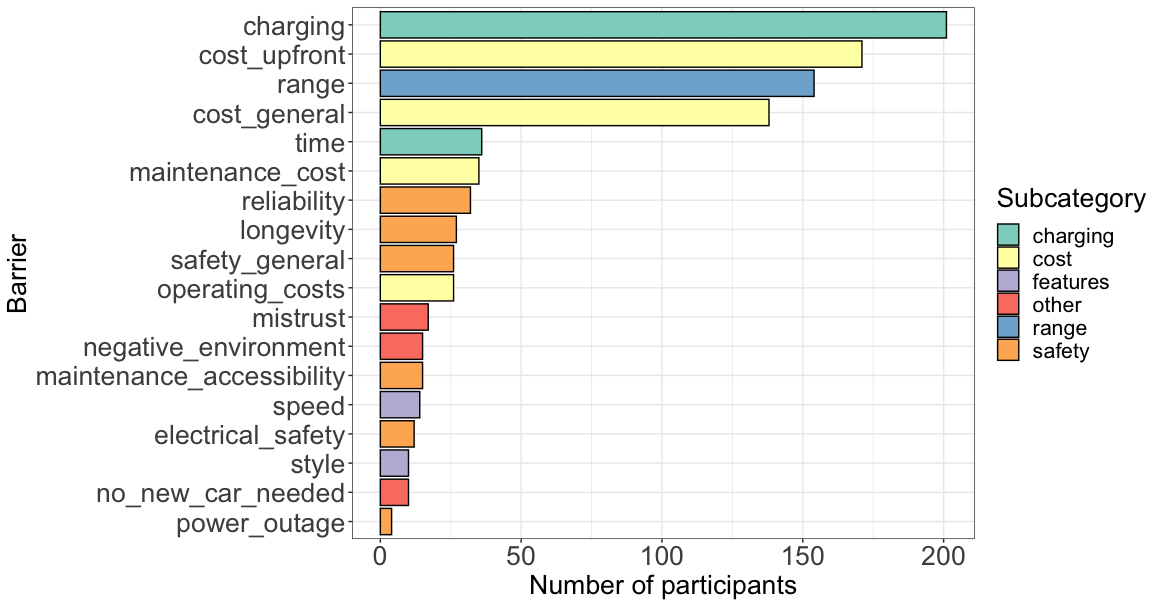}
    \caption{The raw counts of each time a barrier was mentioned by a participant. Note participants could list more than one barrier.}
    \label{fig:barrierPlot}
\end{figure}

An alternative, indirect approach to predicting personalized interventions that might be most effective for a given individual is to instead use elicited factors. Many factors can be assessed per person, thus reducing the number of subjects needed. We hypothesize that factors such as barriers and motivators towards a vehicle engine preference can inform the prediction of interventions. 
Examples of barriers are shown in Figure \ref{fig:barrierPlot}.  Since such factors are elicited from a person, each person can provide information about their set of barriers, enabling collection of information about the best type of intervention for each person. This contrasts with the constraint of asking a person about the effect of only one intervention or a small number of interventions.

\subsection{Barriers Dataset}
We ran a study asking 500 drivers in the United States for demographic information in a multiple-choice format and for their top barriers in a free-response format. The top barriers were then coded into a set of barriers, a subset of which are shown in Figure \ref{fig:barrierPlot}.  We filtered out all barriers that were mentioned fewer than 10 times, except for `power\_outage', which was kept because it was the only barrier mentioned by one person. This produced the list of barriers in Figure \ref{fig:barrierPlot} that we used in our experiments.

The answers by each subject to set of 14 demographic questions were collected as part of the study. The answers were then considered as features for predicting barriers. Many of the features can be intuitively related to barriers. For example, income bracket may influence how likely `cost\_upfront' (e.g., the price of the vehicle) is to be a barrier, and whether a person lives in a house or apartment or whether a person lives in an urban or rural location may influence whether `charging' is a barrier.

The demographic features we used included whether they live in a house/townhouse or apartment, rent, have charging available, live in an urban or rural area, age bracket, education level, members in household, are employed, income bracket, political leaning, gender, prefer gas or BEVs, and whether their driving includes commutes, long distances, or off road.



\subsection{Prediction and Ranking of Barriers}
Since the intent is to present one intervention at a time to a subject, the interventions should be ranked to inform the presentation order, rather than simply classified as to inform whether to present. We envision that the rank of an intervention roughly corresponds to the importance of the barrier that the intervention addresses. Thus, the barriers should also be predicted and ranked for each person, where the predictions are based on a subject's demographics.  There were a total of 18 barriers to be predicted in a multi-label task, since subjects were allowed to list more than one barrier.
We examined two models: (1) simple Multi-Layer Perceptron (MLP) and (2) Support Vector Machine (SVM). 

\subsubsection{MLP} 
We used 3 layers with a logistic activation function and 10 nodes. The output layer uses a softmax nonlinearity, where each of 18 nodes corresponded to a barrier. The MLP performed multi-label classification using a mean squared error loss.

\subsubsection{SVM} With the small dataset size, we also examined the use of an SVM. To handle the multi-label classification task, a one-vs-rest model was used for each of the 18 barriers. To further reduce the effect of the small dataset, we used cross-validation with stratification of the folds for training and evaluating the model. This enabled us to use 95\% of the data for training when there are at least 20 exemplars for a class. When there were fewer than 20 exemplars, the number of partitions was set to the number of exemplars. Both an RBF and a polynomial kernel of order 3 were examined; we report the results for the polynomial kernel, which had slightly better performance. Since the target barriers were highly imbalanced, the weighting of the target and background class was balanced so that the weights were inversely proportional to their frequencies.

\begin{figure}
    \centering
    \includegraphics[width=\columnwidth]{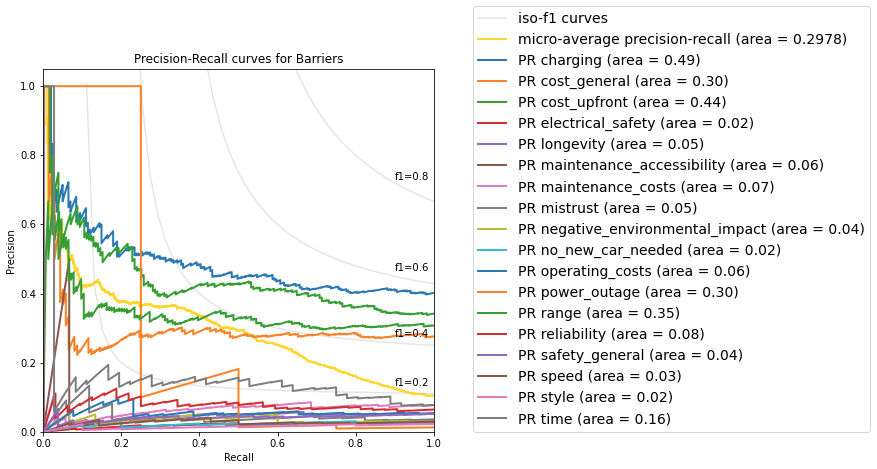}
    \caption{Precision-Recall curves for 18 barriers.}
    \label{fig:PRcurves_barrier}
\end{figure}

\subsection{Barrier Prediction Evaluation}
In the survey to collect our dataset, the subjects listed their top barriers but did not rank them. Thus, the evaluation task is similar to an information retrieval task of identifying and ranking relevant documents for presenting to a user, where documents are only labeled as relevant or not. We used Precision and Recall for evaluation, a metric commonly used to evaluate information retrieval tasks. 

\subsubsection{Ranking Performance}
Figure~\ref{fig:PRcurves_barrier} shows the precision vs recall curves for the SVM model for each of the 18 barriers, as well as the macro-average over all barriers. The overall macro precision-recall (PR) area under the curve is 0.2968. Note that the barriers that were mentioned more frequently tend to have higher PR values, possibly related to the effect when the number of targets available for training is small. 

\begin{figure}
    \centering
    \subfloat[\centering]{\includegraphics[width=.5\linewidth]{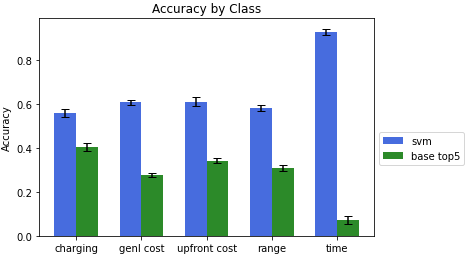}}    
    \subfloat[\centering]{\includegraphics[width=.5\linewidth]{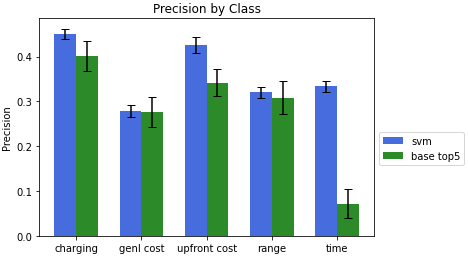}}
    \caption{(a) Accuracy and (b) precision (error bars indicate one standard error) of the SVM model and a baseline of always predicting each of the top 5 most frequent barriers.}
    \label{fig:classification_results}
\end{figure}

The MLP model predicted the same ranking of barriers with the less than 1\% difference in probability values for all individuals. The top-ranked barriers corresponded to the most frequently mentioned barriers. Thus, for a baseline model, we used the most frequently mentioned barriers.  

\subsubsection{Classification Performance}
Since the probability of each barrier varies less than 1\% over the users, we evaluated the performance in a classification task.
The classification performance of always predicting the top-5 barriers compared to the classification performance of the SVM models for the same barriers is shown in Figure~\ref{fig:classification_results}. Precision and accuracy were used. (We did not consider recall or hit rate and false alarm rate since the baseline always predicts each of the top-5, so each of the measures is always 1.) These measures are shown for the SVM and the baseline models in Figure~\ref{fig:classification_results}. The mean accuracy and precision of the SVM model for each of the classes is higher then presenting the top 5. 

\begin{figure}
    \centering
    \includegraphics[width=\columnwidth]{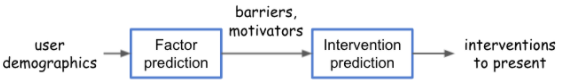}
    \caption{Intervention selection based on predicted barriers.}
    \label{fig:barriers2interventions}
\end{figure}

\subsection{Use of Predicted Barriers for Intervention Selection}
From the predicted barriers, the interventions that best address those barriers can then be identified for presentation to a user, as shown in Figure~\ref{fig:barriers2interventions}.
The interventions are manually defined by experts in Behavioral Science and for rich modalities, also by experts in Human Computer Interaction. If there are multiple interventions for a barrier, a study to assess the effectiveness of each intervention (as measured by preference shift) by presenting one or a few interventions per subject can be conducted, and the effectiveness of the intervention (see the MTurk study described in the next section) used when deciding which intervention to present for a predicted barrier.


\begin{figure*}[h!]
   \centering
   \begin{tabular}{ccc}
    \includegraphics[width=0.3\textwidth]{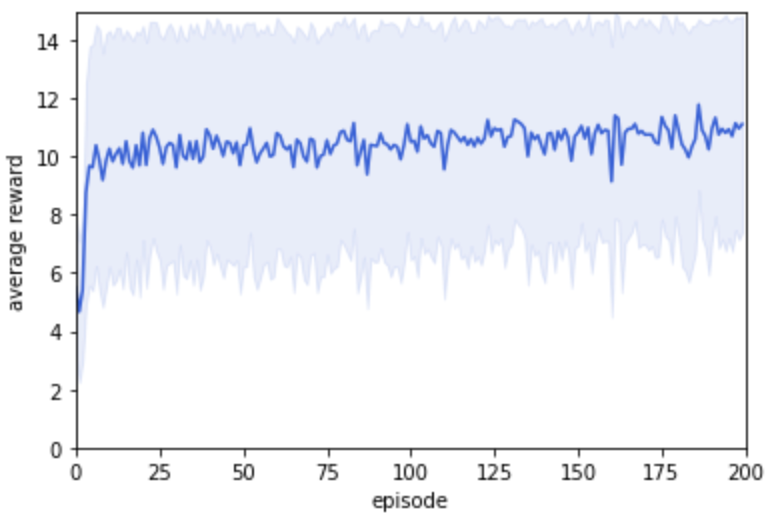} &
    \includegraphics[width=0.3\textwidth]{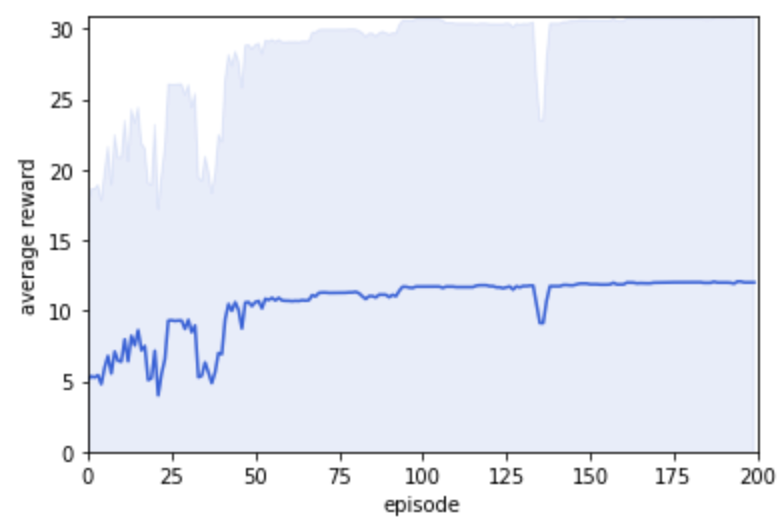} &
    \includegraphics[width=0.3\textwidth]{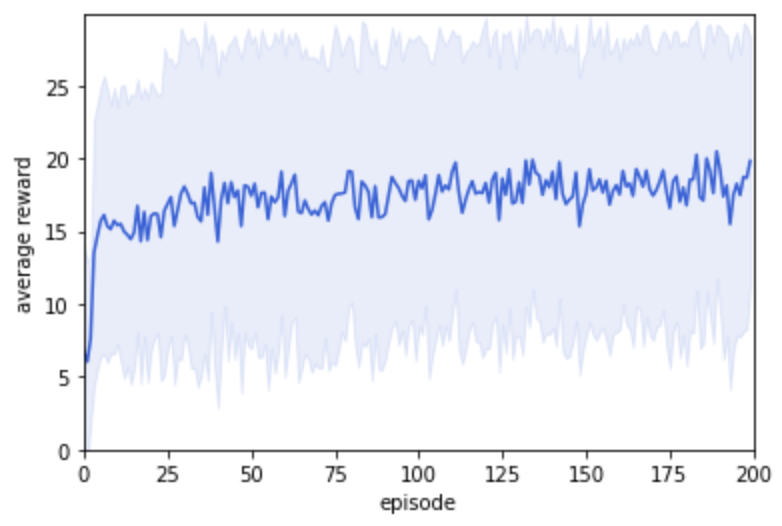} \\
    \textbf{(a)} & \textbf{(b)} & \textbf{(c)} \\
   \end{tabular}
    \caption{(a) Learning curve for the non-demographic, deterministic human model. (b) Learning curve for the non-demographic, stochastic human model. (c) Learning curve for the demographic-aware, deterministic human model. In each case, interactions are bundled in episodes of 100 interventions each, and the curve represents the mean of those 100 interventions. Shaded areas around the average represent standard deviations.}
\label{fig:learningcurves}
\end{figure*}

\section{Recommending Interventions with Reinforcement Learning}\label{sec:rl_interventions_and_survey}
To recommend interventions, we would like to use a reinforcement learning model to learn the best intervention for each user. Since reinforcement learning approaches are data-hungry and participant data is expensive to gather, we decided to run an initial data-gathering survey which would inform simulation models. These simulation models could then be used to pre-train reinforcement learning models, and allow us to select the model that is most likely to perform well in deployment. We experimented with a model that takes account demographics and one that does not.

To gather data for our simulation models,
 4136 MTurk subjects participated in a survey to measure intervention effectiveness. At the beginning of the survey, the subjects answered the demographic questions and indicated their initial preferences for BEVs. Then they were randomly exposed to one of the 35 interventions we designed and then they answered the BEV preference question again. A preference changes score was computed as the difference between pre- and post-intervention answers. For each intervention, we computed a mean intervention effectiveness. The mean intervention effectiveness ranged from $0.275$ to $13.10$, with an
average of $5.52$, and a standard deviation of $2.9$.


\subsection{Non-Demographic Model}
We developed a reinforcement learning model that learns to select the intervention that is most effective, on average, regardless of demographic information. To train this model, we developed two models to simulate human responses to interventions. The first is a deterministic model, in which the preference shift is always equal to the mean preference shift observed for each intervention. The second is a stochastic model, in which the preference shift is modeled by a Gaussian distribution with mean and variance equal to those observed in the data.

The reinforcement learning agent uses an $\epsilon$-soft policy with a SARSA update.  We anneal the exploration parameter $\epsilon$ according to the following schedule, where $t$ is a time variable incremented at every step at which the agent learns.
\begin{equation}
   \small \epsilon(t) = \frac{0.4}{1+(1\times 10^{-5}) t}
\end{equation}
The learning rate is similarly annealed according to the following schedule.
\begin{equation}
   \small \alpha(t) = \frac{10}{1+(1\times 10^{-2}) t}
\end{equation}

Figure \ref{fig:learningcurves}(a) shows the learning curve for the non-demographic scenario in which the simulation model of the human deterministically shifts its preferences with the mean observed preference shift for each intervention. We have bundled the interventions into ``episodes'' of 100 interventions each, even though the human model is stateless, so that previous interventions do not have an effect on future interactions. The solid blue curve represents the mean reward in each episode, and the shaded area represents the standard deviation of reward during the episode.

Figure \ref{fig:learningcurves}(b) shows the learning curve for the non-demographic scenario with the stochastic human model. We observe that the model converges more slowly than for the deterministic case, and also that its performance has much more variance because the human model itself is stochastic.

\subsection{Demographic-Aware Model}
The models without demographics inform a baseline on the effectiveness of interventions that are conducted independent of demographics. Next, we consider a simulation of a survey in which different participants with different demographics are presented. We selected two genders and two age groups that were well-represented in the data, and created a simulation model for a survey that presented randomly selected individuals from those four possible demographic combinations.

Figure \ref{fig:learningcurves}(c) shows the learning curve for the survey experiment. We note that the learning curve has higher variance than that of Figure \ref{fig:learningcurves}(a), but also that the intervention effectiveness is converging to a higher value. This reflects the effectiveness of targeting interventions with the help of demographic information.

In these experiments, we used  two types of demographics and hypothesize performance may improve with additional demographics. However, the improvement would trade off with a requirement for more training episodes to learn the larger space, which we leave for future work.

\section{Discussion and Future Work}
We proposed two approaches to handle the limited amount of human subject data for suggesting interventions to increase electric vehicle preference. We have demonstrated the use of machine learning models to predict barriers to the adoption of electric vehicles from demographic information, as well as to suggest interventions. Our preliminary results demonstrate that better intervention effectiveness can be attained by tailoring the choice of intervention to the demographics of the specific survey participant. In future work, we will validate our trained models on new survey participants, and compare the effectiveness of the interventions suggested by the RL model against interventions recommended by a human expert to address barriers that are predicted for the participant.


\bibliography{bibliography}

\end{document}